\documentclass{article}
\usepackage{spconf,amsmath,graphicx}

\usepackage{enumitem}
\setlist{nosep, leftmargin=14pt}

\usepackage{mwe} 

\title{Efficient brain age prediction from 3D MRI \\volumes using 2D projections}
\name{Johan Jönemo$^{\: a,d}$, Muhammad Usman Akbar$^{\: a,d}$, Robin Kämpe$^{\: c,d}$, J. Paul Hamilton$^{\: c,d}$, Anders Eklund$^{\: a,b,d}$}

\address{$^a$Division of Medical Informatics, Department of Biomedical Engineering\\ $^b$Division of Statistics \& Machine Learning, Department of Computer and Information Science\\ $^c$Center for Social and Affective Neuroscience, Department of Biomedical and Clinical Sciences \\ $^d$Center for Medical Image Science and Visualization (CMIV) \\Link\"{o}ping University, Sweden}

\begin{document}
%
\maketitle
\begin{abstract}
Using 3D CNNs on high resolution medical volumes is very computationally demanding, especially for large datasets like the UK Biobank which aims to scan 100,000 subjects. Here we demonstrate that using 2D CNNs on a few 2D projections (representing mean and standard deviation across axial, sagittal and coronal slices) of the 3D volumes leads to reasonable test accuracy when predicting the age from brain volumes. Using our approach, one training epoch with 20,324 subjects takes 20 - 50 seconds using a single GPU, which is two orders of magnitude faster than a small 3D CNN. These results are important for researchers who do not have access to expensive GPU hardware for 3D CNNs.
\end{abstract}
\section{Introduction}
\label{sec:intro}

Predicting brain age from MRI volumes using deep learning has become a popular research topic recently~\cite{bjork2018ct,cole2017predicting,jonsson2019brain,peng2021accurate,bellantuono2021predicting,gupta2021improved,dinsdale2021learning}. If there is a large difference between the predicted brain age and the biological age of a patient, one can suspect that some disease is present and the difference is therefore an important biomarker. Virtually all of the previous works have used 3D CNNs to predict the brain age, or trained 2D CNNs on slices and then combined all the slice predictions to a prediction for the entire volume~\cite{gupta2021improved}. Since 3D CNNs are computationally demanding and require lots of GPU memory, we therefore propose to instead use 2D projections of the 3D volumes. Compared to previous approaches that use 2D CNNs on volume data~\cite{gupta2021improved}, we only use 1 - 6 images per patient (compared to using all 100 - 300 slices in a volume).

Using 2D CNNs has many benefits compared to 3D CNNs. For example, 2D CNNs can use cheaper hardware (important for low income countries), can use networks pre-trained on ImageNet or RadImageNet~\cite{radimagenet} (there are very few pre-trained 3D CNNs) and in general benefit from the more mature and better optimised 2D CNN ecosystem.
They can also have fewer parameters (which for example makes federated learning easier due to less communication). Furthermore, due to the faster training it is much easier to tune the hyperparameters.

Langner et al.~\cite{langner} demonstrated that 2D projections of full body MRI volumes can be used to train 2D CNNs to predict different measures like age. Since brain volumes contain less anatomical variation compared to full body volumes, it is not clear if the same approach is well suited for brain volumes. Furthermore, Langner et al. only used mean intensity projections, while we also use the standard deviation projections (to better capture the variation between slices). \\

\begin{figure}[htb]
\begin{minipage}[b]{1.0\linewidth}
  \centering
  \includegraphics[scale = 0.29]{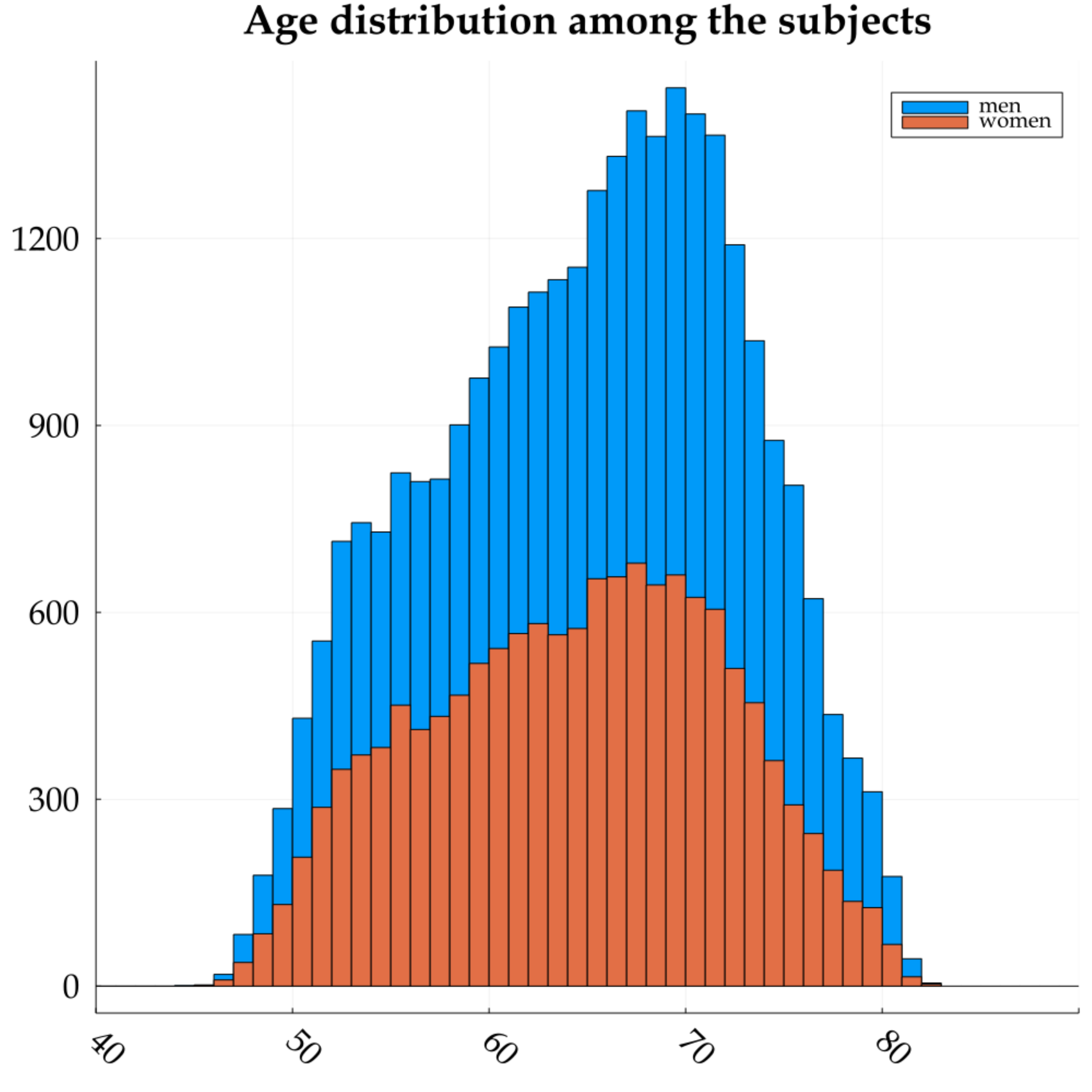}
\caption{Age distribution for the 29,034 subjects used in this work. The individual bars are further divided to reflect the proportion of each gender within that age group}
\end{minipage}
\label{fig:age}
\end{figure}

\begin{figure}[htb]
\begin{minipage}[b]{0.99\linewidth}
  \centering
  \includegraphics[scale = 0.32]{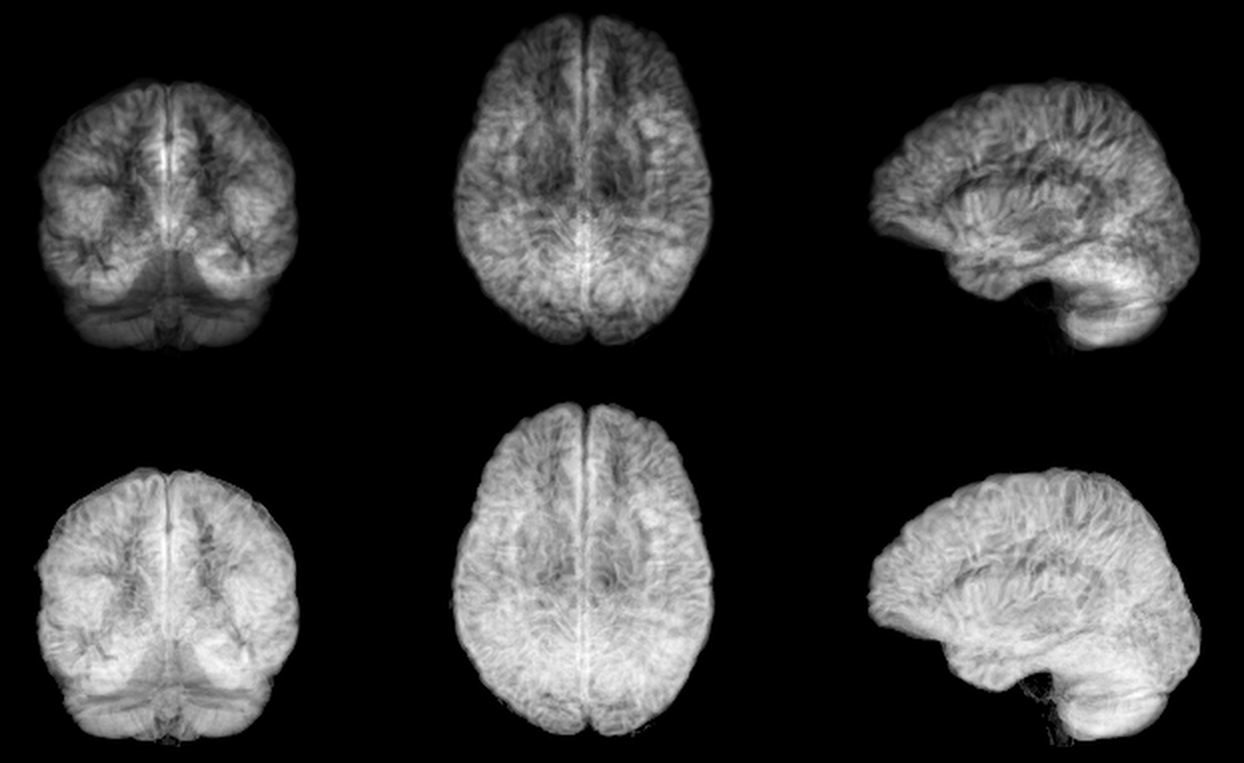}
\caption{\textbf{Top:} mean grey matter likelihood projections on coronal, axial and sagittal planes, for one subject. \textbf{Bottom:} standard deviation grey matter likelihood projections on coronal, axial and sagittal planes, for the same subject. 
}
\end{minipage}
\label{fig:data}
\end{figure}

\vspace{-0.5cm}
\section{Data}
\vspace{-0.1cm}
The experiments in this paper are based on 29,034 T1-weighted brain volumes from UK Biobank~\cite{sudlow2015uk,alfaro2018image,littlejohns2020uk}. The subjects were divided into 20,324 for training, 4,356 for validation and 4,355 for testing. FSL FAST~\cite{zhang2001segmentation} was used for each skull-stripped volume, to obtain maps of gray matter. These gray matter volumes were zeropadded, symmetrically, to match the largest grid (matrix-size), resulting in volumes of 256 x 256 x 208 voxels. Each volume was then projected into six 2D images, which represent the mean and standard deviation across axial, sagittal and coronal slices (for one subject at a time). The age range is 44 - 82 years with a resolution of 1 year, see Figure~\ref{fig:age} for the age distribution. See Figure 2 for the six projections of one subject. The original dataset is about 1.5 TB as 32 bit floats.

\begin{figure}[htb]
\begin{minipage}[b]{1.0\linewidth}
  \centering
  \includegraphics[scale = 0.14]{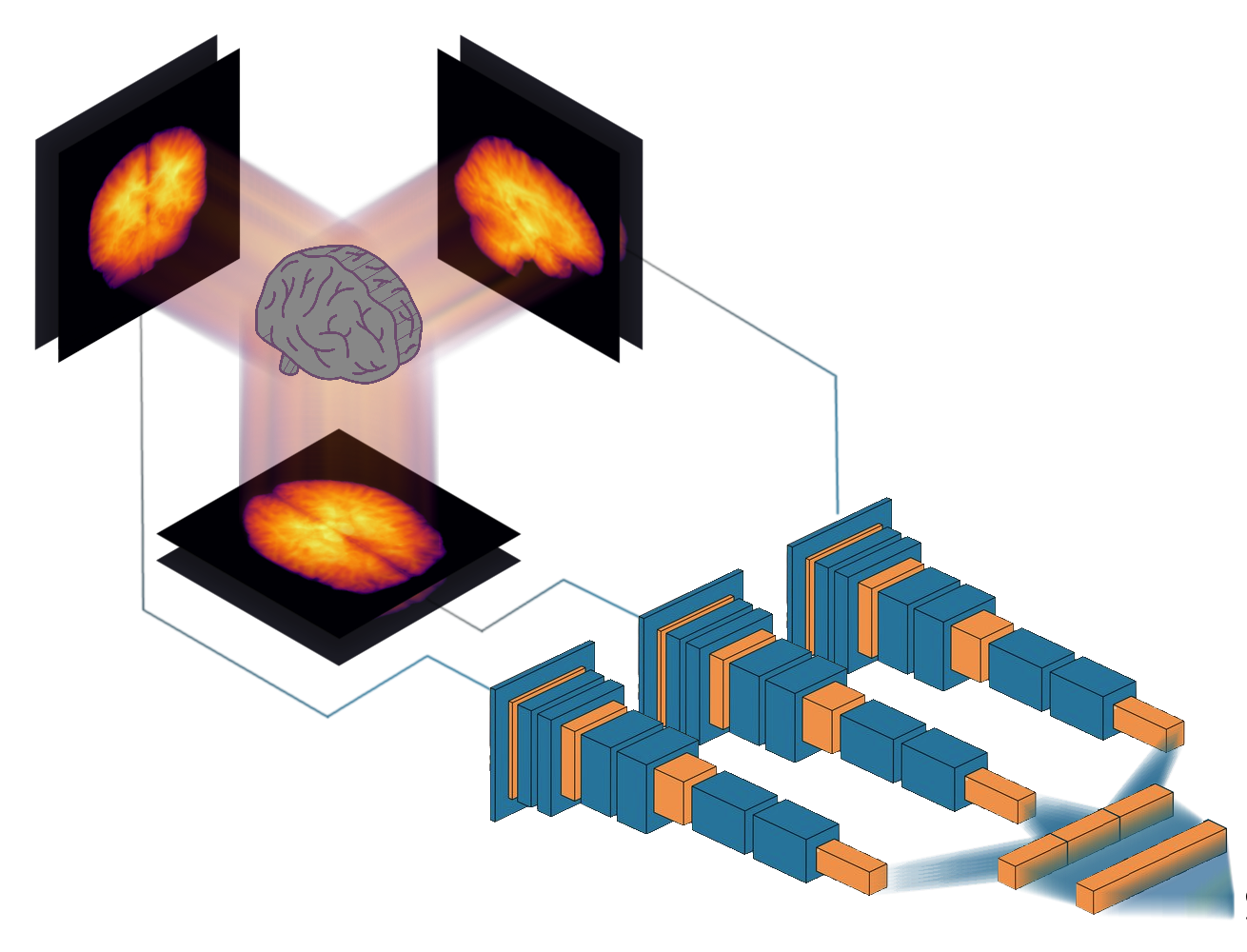}
  \caption{Our proposed approach to obtain efficient brain age prediction using 2D projections of 3D volumes. Each volume is summarized as six 2D images, which represent the mean and standard deviation across axial, sagittal and coronal slices. These 2D images are then fed into three 2D CNNs, and the resulting feature vectors are concatenated and fed into a fully connected layer to predict the brain age. }\medskip
\end{minipage}
\label{fig:network}
\end{figure}

\vspace{-0.3cm}
\section{Methods}

In this work we have implemented a set of 2D CNNs using the Julia programming language (version 1.6.4)~\cite{bezanson2017julia} and the Flux machine learning framework (version 0.12.8)~\cite{innes2018flux}, wherein the aforementioned projections - typically with two channels each - are fed into their respective stack of convolutional and auxiliary layers (see Figure 3). Instead of training a single multi-channel CNN, three separate CNNs are trained as the important features for e.g. sagittal images may be different from the important features for e.g. axial images. Each CNN produces 256 features which are concatenated and fed into a fully connected layer ending in one node with linear output.

The models tested had 13 convolutional layers for each projection (axial, coronal or sagittal). The convolutional stacks had 4 filters in the first layer which then progressed as the resolution was reduced to 256 filters as mentioned earlier. The models had from a little more than 0.8 million to 2 million trainable parameters.

The training was done using mean squared error as a loss function. Batch normalization and dropout regularization (probability 0.2) was used after every second convolutional layer, or between the dense layers (probability 0.3 or 0.5). Optimization was done using the Adam optimizer, with a learning rate of 0.003. Training was always performed for 400 epochs, and the weights were saved every time the validation loss decreased. Furthermore, the training was also performed where the weights of the three 2D CNNs were fixed to be the same (called iso).

 Data augmentation was tentatively explored using the Augmentor module~\cite{bloice2017augmentor}, wherein an augmentation pipeline was constructed. The augmented data set consisted of the unaugmented set concatenated with three copies that had been passed through a pipeline of small random pertubations in the form of scaling, shearing, rotation and elastic deformation. This set was randomly shuffled for each epoch of training. As of yet the code has not successfully been made to work with on-the-fly augmentation, nor have we been able to utilise GPUs for these calculations. 
 

Training the networks was performed using a Nvidia RTX 8000 graphics card with 48 GB of memory. A major benefit of our approach is that all the training images fit in GPU memory (when augmentation is not used), making the training substantially faster since the images do not need to be streamed from CPU memory or from the hard drive. One epoch of training with 6 projections from 20,324 subjects takes 20 - 50 seconds (which can be compared to 1 hour for a 3D CNN trained with 12,949 subjects~\cite{peng2021accurate}). Our code is available at https://github.com/emojjon/brain-projection-age


\begin{table*}[htb] \large
    \centering
    \begin{minipage}{\textwidth}
    \begin{tabular}{c|c|c|c|c}
        \textbf{Paper} &  \textbf{N subjects} &  \textbf{Test accuracy} &  \textbf{Parameters} &  \textbf{Training time}  \\ 
      \hline

 Cole et al~\cite{cole2017predicting}, $2017$  &  1601 &  4.16 MAE &  889,960 &  72 - 332 h  \\
 Jonsson et al~\cite{jonsson2019brain}, 2019  &  809  &  3.39 MAE &  ?  &  48 h  \\

 Peng et al.~\cite{peng2021accurate}, 2021 &  12,949  &   2.14 MAE &  3 million &  130 h\\
 Bellantuono et al~\cite{bellantuono2021predicting}, 2021  &  800 ? &  2.19 MAE  &  ? &  ? \\
 Gupta et al~\cite{gupta2021improved}, 2021  &  7,312  &  2.82 MAE  &   998,625 &  ?\\
 Dinsdale et al~\cite{dinsdale2021learning}, 2021  &  12,802  &   2.90 MAE &   ? &  ?\\


\hline
Dropout between conv  &  &  &  &\\
0.2 dropout rate & & & & \\


   Ours, 3 mean channels  &  20,324  & 3.55 (4.49)   & 2,009,261 & 22 min (3 h 53 min)  \\
  Ours, 3 std channels   &  20,324  & 3.51 (4.43) &  2,009,261 & 24 min (3 h 30 min)  \\
  Ours, all 6 channels   &  20,324  &  3.53 (4.44)  & 2,009,369 & 24 min (3 h 26 min) \\
  Ours, all 6 channels, iso &  20,324  &  3.46 (4.38)   &  827,841 & 25 min (4 h 36 min) \\

 & & & & \\

\hline
Dropout between dense &  &  &  &\\
0.3 dropout rate & & & & \\


  Ours, 3 mean channels  &  20,324  & 3.70 (4.66)   & 2,009,261 & 22 min (3 h 12 min)  \\
  Ours, 3 std channels   &  20,324  & 3.67 (4.62) &  2,009,261 & 27 min (4 h 27 min)  \\
  Ours, all 6 channels   &  20,324  &  3.56 (4.47)  & 2,009,369 & 27 min (3 h 32 min) \\
  Ours, all 6 channels, iso &  20,324  &  3.63 (4.56)   &  827,841 & 28 min (4 h 23 min) \\

 & & & & \\

\hline
Dropout between conv &  &  &  &\\
0.2 dropout rate & & & & \\
trained with augmentation& & & & \\


  Ours, 3 mean channels  &  20,324\footnotemark[1]  & 3.44  (4.31)   & 2,009,261 & $>$ 3 days\footnotemark[2]  \\
  Ours, 3 std channels   &  20,324\footnotemark[1]  & 3.40  (4.33) &  2,009,261 & $>$ 3 days\footnotemark[2]  \\
  Ours, all 6 channels   &  20,324\footnotemark[1]  & 3.47 (4.40)  & 2,009,369 & $>$ 3 days\footnotemark[2] \\
  Ours, all 6 channels, iso &  20,324\footnotemark[1]  &  3.85 (4.80) & 827,841 & $>$ 3 days\footnotemark[2]  \\
 & & & & \\

\hline

 

 

    \end{tabular}
    \caption{Comparison of our 2D projection approach and previous publications on brain age prediction (using 3D CNNs), regarding number of training subjects (N), brain age test accuracy (mean absolute error (MAE) in years, RMSE in parenthesis) and training time. Iso here refers to that the 3 parallel 2D CNNs (for axial, sagittal and coronal projections) are forced to use the same weights. Even though several publications use the UK biobank data, a direct comparison of the test accuracy is not possible as different test sets, in terms of size and the specific subjects, were used. The available training times were rescaled to a single GPU, if multi-GPU training was mentioned. The training time for our approach is presented for early stopping, and for the full 400 epochs in parenthesis.}
    \label{tab:res}
    \renewcommand{\thempfootnote}{\arabic{mpfootnote}}
    \footnotetext[1]{The model is trained with an augmented set of 20,324 + 60,972 = 81,296 pseudo subjects, but all are derived from the original 20,324 subjects}
    \footnotetext[2]{This was a preliminary exploration of whether augmentation was motivated. For more competitive speeds some optimisation is needed.}
    \end{minipage}
\end{table*}

\begin{figure}[]
\begin{minipage}[b]{0.99\linewidth}
  \centering
  \centerline{\includegraphics[scale=0.30]{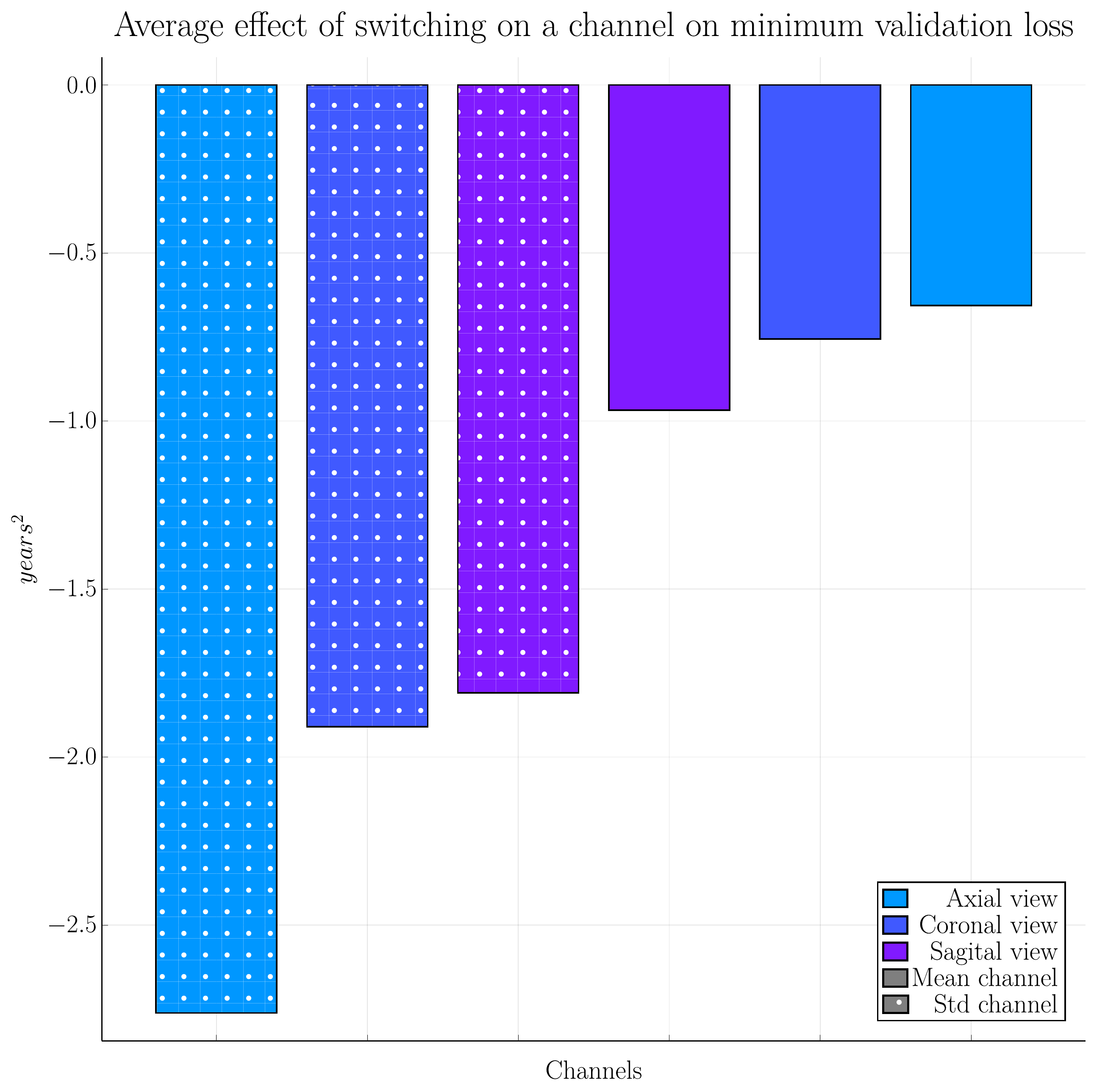}}
  \caption{The effect - in the preliminary study on raw intensities - of adding additional channels on the prediction accuracy, averaged over 128 trainings when using different combinations of input channels (64 different input combinations for two different learning rates). Adding the standard deviation images from the different views have the largest effects and the mean images the smallest.  }\medskip
\end{minipage}
\label{fig:channels}
\end{figure}

\vspace{-0.25cm}
\section{Results}

Table~\ref{tab:res} shows the test prediction accuracy and training time for previously published papers (using 3D CNNs, or 2D CNNs on all slices) and our approach using 2D projections. While several papers used the UK Biobank dataset, the test sets are different which makes a direct comparison of the test accuracy difficult (we would need to implement and train all other networks on our specific data). Our approach is substantially faster compared to the previously published papers, even though we are using the largest training set, while our test accuracy is worse. Using the standard deviation to produce 2D projections leads to a slightly higher accuracy, compared to using the mean across slices. Using both mean and standard deviation projections sometimes provides a small improvement, compared to only using the standard  deviation. Forcing the three 2D CNNs to use the same weights (referred to as iso) sometimes leads to a higher accuracy, compared to using three independent CNNs. Data augmentation helps to further improve the accuracy, but is currently much slower.

In a preliminary study we trained the 2D CNNs repeat-
edly with 1 - 6 input projections from the original intensity (the results largely follow the same pattern as grey matter likelihood but with
slightly lower accuracy), to see which projections that are
most important for the network, resulting in a total of 64 combinations. This was repeated for two learning rates, for a total of 128 trainings. Figure 4 shows the decrease in loss when adding each channel, averaged over said trainings. Clearly, the standard deviation projections are more informative compared to the mean intensity projections. 

\vspace{-0.3cm}
\section{Discussion}

Our results show that our 2D projection approach is substantially faster compared to previous 3D approaches, although several papers do not report the training time. The speedup will in our case not be as large for GPUs with smaller memory, as it is then not possible to put all the training images in GPU memory (for a preliminary test on a 11 GB card, the training took 3-4 times longer but this can probably be further optimized). Nevertheless, the possibility to use cheaper hardware is important for many researchers. Our test accuracy is, on the other hand, a bit worse compared to 3D CNNs, but our work should rather be seen as a proof of concept. Compared to recent 2D CNNs, our network is rather shallow with 13 convolutional layers, which may explain the lower accuracy. It would be interesting to instead use 2D CNNs pre-trained on ImageNet or RadImageNet~\cite{radimagenet} as a starting point, instead of training from scratch. However, this option is currently more difficult in Flux compared to other machine learning frameworks. 

Langner et al.~\cite{langner}, who used 2D projections of full body MRI scans (not including the head), obtained a mean absolute error of 2.49 years when training with 23,120 subjects from UK Biobank (training the network took about 8 hours). It is difficult to determine if the higher accuracy compared to our work is due to using a VGG16 architecture (pre-trained on ImageNet), or due to the fact that full body scans contain more information regarding a person's age, or that the full body scans in UK Biobank contain separate images representing fat and water. No comparison with a 3D CNN is included in their work.

In future work we plan to investigate the effect of adding additional images (channels) which represent the third and fourth moment (skew and kurtosis) across slices, since the results indicate that the standard deviation images are more informative compared to the mean intensity images. Another idea is to use principal component analysis (PCA) across each direction, to instead use eigen slices that represent most of the variance. As can be seen in Table~\ref{tab:res}, adding more channels will not substantially increase the training time as a higher number of input channels will only affect the first layer of each 2D CNN. This is different from adding more training images to a 2D CNN using each slice in a volume independently, where the training time will increase more or less linearly with more images.

\vspace{-0.4cm}
 \section{Ethics and Acknowledgments}
\vspace{-0.1cm}
This research study was ethically approved according to the Swedish ethical review authority, application number 2017/17-31. This research was supported by the ITEA / VINNOVA project ASSIST (Automation, Surgery Support and Intuitive 3D visualization to optimize workflow in IGT SysTems, 2021-01954), LiU cancer and the Åke Wiberg foundation.


%
%
%


\clearpage

%



\bibliographystyle{IEEEbib}
\bibliography{refs}

\end{document}